\documentclass[amsmath,amssymb]{revtex4}

\begin{document}

\title{Cumulant-based calculations of the correlation energy in a molecule}
%\title{Кумулянтные методы вычисления корреляционной энергии молекулы}
%\tnotetext[mytitlenote]{Fully documented templates are available in the elsarticle package on.}

\author{A. K. Zhuravlev}
\email {zhuravlev@imp.uran.ru} %\affiliation{Institute of Metal Physics, 620108 Ekaterinburg, Russia}
\affiliation{M.N. Miheev Institute of Metal Physics of Ural Branch of Russian Academy of Sciences, 620108 Ekaterinburg,
Russia}
%\address{M.N. Miheev Institute of Metal Physics of Ural Branch of Russian Academy of Sciences, 620990 Ekaterinburg, Russia}

\begin{abstract}
The problem of constructing a guaranteed convergent sequence of corrections to the Hartree--Fock ground state energy of a
molecule without storing the many-electron wave function
%((in computer memory))
is considered. Several methods based on cumulants
are considered and it is shown that such a sequence is obtained by Lanczos tridiagonalization, in which the elements of the
tridiagonal matrix are calculated through cumulants.
\end{abstract}

%M.N. Miheev Institute of Metal Physics of Ural Branch of Russian Academy of Sciences
%\maketitle
%\date{\today}


\maketitle

%The problem of strongly correlated quantum many-body systems is one of the most complicated in theoretical physics. With
%the exception of a few of simplified models this problem cannot be solved analytically, so that one must resort to
%numerical methods. But here a researcher is faced with serious difficulties. For example, exact diagonalization runs into
%the exponential growth of the Hilbert space dimension with increasing size of the system and therefore is limited to small
%clusters, even when using the Lanczos algorithm~\cite{Dagotto1994}. A more sophisticated Density-Matrix Renormalization
%Group technique with high-energy states truncation~\cite{DMRG} gives excellent results for the ground state energy of
%one-dimensional Fermi systems, but has its own limitation when applied to two- and three-dimensional
%cases~\cite{Liang_Pang}. Quantum Monte Carlo method \cite{QMC} can potentially handle larger systems. However, the method
%works poorly at low temperatures for fermion systems because of the so called ``minus-sign'' problem~\cite{QMC_minus-sign}.

The calculation of the correlation energy, defined as the difference between the exact energy of the electronic system and
the approximate energy calculated by Hartree--Fock (HF) method, is a classic problem in quantum chemistry.
%has been является классической проблемой квантовой химии.
%a subject of an enormous amount of theoretical study during last 30 years. exact diagonalization runs into the exponential
%growth of the Hilbert space dimension with increasing size of the system and therefore is limited to small clusters

%Метод Хартри--Фока (ХФ) давно и широко используется в квантовой химии. %Корреляционной энергией называется разность между энергией
%основного состояния в приближении Хартри--Фока и истинной энергией основного состояния системы. Правильное вычисление этой
%величины является одной из важнейших задач квантовой химии, ибо
%Однако разница между хартрифоковской и истинной энергиями основного состояния может оказаться порядка энергии связи атомов
%в молекуле. Поэтому ошибка, даваемая методом Хартри--Фока, может стать фатальной при принятии решения, пойдет реакция
%синтеза атомов в молекулу или нет.
%Как правило,
%Наиболее прямолинейный способ определения корреляционной энергии - (ФЦИ ) -  приводит к экспоненциальному росту требований
%к компьютерной памяти с ростом размера молекулы
The most straightforward way of the correlation energy calculation -- full
configuration interaction method (FCI) -- leads to an exponential increase in the requirements for computer memory with an
increase in the size of the molecule.
%Наиболее прямым способом получить систематические поправки к хартрифоковской энергии
%получаются за счет увеличения числа взятых в рассмотрение многоэлектронных конфигураций, что приводит к экспоненциальному
%росту требуемой для расчета памяти компьютера с увеличением размера молекулы (числа молекулярных орбиталей).
%Поэтому с помощью
%полноконфигурационных методов можно рассчитать лишь весьма небольшую молекулу.
%Поэтому актуальной задачей является
%построение метода, способного обойтись без такого роста.
The M{\o}ller--Plesset (MP) perturbation theory \cite{MP_1934} can do without such growth. To use this method, only Coulomb
integrals and Hartree--Fock energy levels are needed \cite{Cremer_review2011}.
%For the use of which only Coulomb integrals and Hartree-Fock energy levels are needed.
%Без такого роста может обойтись теория возмущений, для использования которой нужны только кулоновские интегралы и хартри-фоковские энергетические уровни.
%\cite{MP_1934,Cremer_review2011}, для вычисление членов которого достаточно только кулоновских интегралов и энергетических
%уровней, вычисленных Хартри-Фоком.
However, the resulting series may be divergent
%Однако получающийся при этом ряд МОЖЕТ ОКАЗАТЬСЯ расходящимся usually diverge~
\cite{Olsen1996,Olsen2000}.

%Существуют методы, основанные на функционале плотности и методы прямого решения исходного уравнения Шредингера. Последние
%наталкиваются на проблему экспоненциального роста размера гильбертова пространства с увеличением размеров молекулы.

%Метод связных кластеров сводится к решению систем уравнений для коэффициентов........ так же является затратным по памяти.

%Хотелось бы иметь процедуру получения систематических поправок к хартри--фоковскому решению, не сильно превышающую по
%затратам (в первую очередь памяти) сам метод Хартри-Фока. Это, конечно, метод М{\o}ллера-Прессета
%\cite{MP_1934,Cremer_review2011}, который фактически является не чем иным, как квантовомеханической теорией возмущений
%Релея-Шредингера. И как известно, ряд этой теории возмущений вполне может оказаться расходящимся.
%На практике это выглядит так, что третий порядок MP может оказаться отстоящим дальше от точного решения, чем второй.

%As an alternative, a certain interest in the construction of regular expansions still exists \cite{Oitmaa}. The attractive
%feature of such alternative is the relative simplicity of calculation of terms in the expansion. Unfortunately, the series
%expansions in powers of the coupling constants МОЖЕТ ОКАЗАТЬСЯ usually diverge~\cite{Olsen1996,Olsen2000}.

%However, there are regular expansion methods which are not reduced to power expansion in coupling constant.

%Можно ли построить с затратами, сравнимыми с MP, гарантированно сходящийся метод?
Is it possible to build a guaranteed convergent method that requires comparable to MP? In the presented work, we will
consider methods related to the calculation of cumulants, and give an affirmative answer to the question asked.

%В данной работе мы рассмотрим методы, связанные с вычислением кумулянтов, и дадим утвердительный ответ на заданный вопрос.

%Мы рассмотрим
%методы, основанные на кумулянтах, так как кумулянты могут быть вычислены без явного использования многоэлектронного
%вектора, так же как и члены ряда МР. И покажем, что существует утвердительный ответ на вышепоставленный вопрос.
\section{Methods}
\subsection{Cumulant $t$-expansion} %Кумулянтное $t$-разложение} %. Calculation of the limit $E(t\rightarrow\infty)$}

%Мы рассмотрим два способа вычисления  (connected moments).
First, consider a method for finding the energy of the ground state, called the cumulant $t$-expansion
\cite{HornWeinstein}.
% First способ вычисления корреляционной энергии, используя кумулянты is the so called $t$-expansion~\cite{HornWeinstein}, which we
%describe briefly in what follows.
Given a Hamiltonian $\hat{H}$ and an initial state $|\phi_0\rangle$, let us define the
moments
\begin{equation}
\mu_m = \langle\phi_0| \hat{H}^m |\phi_0\rangle \label{eq:mu_n}
\end{equation}
($|\phi_0\rangle$ is normalized to unity) and introduce auxiliary function
\begin{equation}
E(t) = \frac{\langle\phi_0| \hat{H}e^{-\hat{H}t} |\phi_0\rangle}{\langle\phi_0| e^{-\hat{H}t} |\phi_0\rangle}
\label{eq:E_t_first}
\end{equation}
which can be written as a power series in the parameter~$t$:
\begin{equation}
E(t) = \sum_{m=0}^\infty \frac{I_{m+1}}{m!}(-t)^m \ ,
\end{equation}
where
\begin{equation}
I_{m+1} = \mu_{m+1} - \sum_{p=0}^{m-1} {m \choose p} I_{p+1} \mu_{m-p} \label{eq:def_cum}
\end{equation}
are the cumulants\cite{Smith} (note that in \cite{Cioslowski,Stubbins} the values $I_m$ were named ``connected moments'').
Then
\begin{equation}
E_0 = \lim_{t\rightarrow\infty} E(t) \label{eq:E0lim}
\end{equation}
is the minimal eigenvalue of the Schr\"{o}dinger equation
\begin{equation}
\hat{H}|\psi_0\rangle = E_0 |\psi_0\rangle \label{eq:Hpsi}
\end{equation}
provided that $\langle \psi_0 |\phi_0\rangle \neq 0$ (see \cite{HornWeinstein} for proof).

%This is one of the main goals of the presented work. In addition, various methods for estimating the ground state energy
%from the calculated cumulants are presented and compared with the full configuration solution for a small molecule.

%\section{Вычисление кумулянтов}
%\section{Вычисление энергии основного состояния}

%Рассмотрим теперь способы вычисления энергии основного состояния $E_0$ по нескольким вычисленным кумулянтам.

%Для того, чтобы вычислить предел (\ref{eq:E0lim}) мы должны знать все кумулянты, что невозможно для сколь-либо реалистичной
%системы. The articles \cite{HornWeinstein,Cioslowski,Stubbins} proposed
Several ways have been proposed to calculate the limit
%Было предложено some ways to calculate the limit
(\ref{eq:E0lim}) using the first few known cumulants \cite{Stubbins}.
%using несколько первых известных кумулянтов
%Кратко рассмотрим два наиболее популярных.
%Somewhat later
In the papers \cite{Cioslowski,Knowles1987}, it was proposed to seek $E(t)$ as a sum of decaying exponentials, which leads
to the following sequence of approximations for the ground state energy, called the connected moments expansion (CMX):
%\begin{equation}
%E^{CMX}_0 = I_1 - \frac{I_2^2}{I_3} - \frac{1}{I_3}\frac{(I_2I_4-I_3^2)^2}{I_3I_5-I_4^2} - \dots \label{eq:CMX}
%\end{equation}
\begin{equation}
    E^{CMX(n)}_0 = I_1 - \begin{pmatrix}I_2 & \cdots & I_n\end{pmatrix}
    \begin{pmatrix}
     I_3     &  \cdots     &  I_{n+1}   \\
     \vdots  &  \ddots     &  \vdots    \\
     I_{n+1} &  \cdots     &  I_{2n-1}
    \end{pmatrix}^{-1}
    \begin{pmatrix}I_2\\ \vdots \\ I_n\end{pmatrix}\ .
  \label{eq:CMX}
  \end{equation}

The CMX method was successfully tested on the problems of calculating the ground state energy of anharmonic oscillator
\cite{Cios1987} and a hydrogen molecule \cite{Cioslowski}. However, problems arose when trying to apply this method to
many-electron lattice models: for some values of the model parameters, expression (\ref{eq:CMX}) became singular
\cite{MassanoBowenMancini1989,ManciniPrieMassano1991,LeeLo1993}.

In \cite{HornWeinstein}, the so-called $D$-Pad\'{e} method was used, in which the Pad\'{e} approximation is applied for the
derivative of the function $E(t)$.
%В работе \cite{HornWeinstein} был применен так называемый $D$-Pad\'{e} method, в котором perform ??? the Pad\'{e}
%approximation for the derivative of the function $E(t)$.
In this case, the first two approximations to the sought ground state energy $E_0$ can be obtained from the expression
%The Figure \ref{Fig_1Dpade} shows the results of several methods of calculating the limit (\ref{eq:E0lim}) when the first
%seven cumulants are known: connected-moments expansion (CMX) by Cioslowski\cite{Cioslowski}, inversion method by
%Stubbins\cite{Stubbins}, diagonal [3/3] Pad\'{e} approximant for $E(t)$ and [0/5]
%$D$-Pad\'{e} approximations used in the
%pioneering work \cite{HornWeinstein}.
%The $D$-Pad\'{e} method рекомендуется в книге \cite{Oitmaa} p.260 yields much more
%accurate results than others and gives a satisfactory results for the model under consideration in the entire range of parameters.
% perform the Pad\'{e} approximation for the derivative of the function $E(t)$ \cite{HornWeinstein}.
\begin{eqnarray}
E^{DP[0/M]}_0 &=& I_1 - \int_0^\infty \frac{I_2}{Q_M(t)}dt , \ \ \textrm{where} \\
Q_2(t) &=& 1 + \frac{I_3}{I_2}t + \frac{2I_3^2-I_2I_4}{2I_2^2}t^2 \ , \nonumber \\
Q_3(t) &=& 1 + \frac{I_3}{I_2}t + \frac{2I_3^2-I_2I_4}{2I_2^2}t^2 +\frac{6I_3^3 - 6I_2I_3I_4 + I_2^2I_5}{6I_2^3}t^3
\nonumber
\end{eqnarray}
(for a more detailed description, see \cite{Zhuravlev2016,Zhuravlev2020}). However, it was shown in \cite{Zhuravlev2020}
that the $D$-Pad\'{e} method is unsatisfactory in the whole range of parameters of the many-electron lattice Hubbard model.
%Однако в работе \cite{Zhuravlev2020} было показано, что Pad\'{e} method неудовлетворителен в целой области параметров
%решеточной модели Хаббарда.

Thus, the listed methods do not provide a guaranteed convergent sequence of approximations for the ground state energy.
%Таким образом, перечисленные методы не дают универсально ??? гарантированно сходящуюся последовательность приближений для
%энергии основного состояния.

\subsection{Cumulant Lanczos tridiagonalization} %Полуаналитический метод Ланцоша (

%Алгоритм (метод) Ланцоша
Lanczos tridiagonalization (LT) is well known
%хорошо известен расчетчикам
in computational many-electron physics \cite{Dagotto1994}.
% многоэлектронных задач как существенно непертурбативный подход к физическим проблемам с сильным взаимодействием.
%В этом методе генерируется последовательность
%ортонормированных состояний $ \{ |\psi_{n}\rangle \}_{n=1,2..} $ и коэффициенты Ланцоша $ \{ \alpha_n \}_{n=0,1..}, \{
%\beta_n \}_{n=1,2..} $, стартуя с затравочного состояния $ |\phi_{0}\rangle $ c помощью следующего рекуррентного
%соотношения:
%The Lanczos algorithm or method has been of interest to physicists because it is an essentially non-perturbative approach
%to physical problems with strong coupling, such as occur in the extensive many-body systems of condensed matter physics.
In this method the Hamiltonian is used to generate a sequence of orthonormal states $\{ |\phi_{n}\rangle \}_{n=1,2..}$ and
Lanczos coefficients $ \{ \alpha_n \}_{n=0,1..}, \{ \beta_n \}_{n=1,2..} $, from a suitably chosen trial state
$|\phi_{0}\rangle$ through the following recurrence
  \begin{equation}
    |\phi_{n+1}\rangle =
          {1\over \beta_{n+1}}
    \left[ (\hat{H} - \alpha_{n}) |\phi_{n}\rangle
                 - \beta_{n} |\phi_{n-1}\rangle \right] \ ,
  \label{eq:def-F}
  \end{equation}
(where $\alpha_n = \langle\phi_n| \hat{H} |\phi_n\rangle$, and $\beta_{n+1}$ -- normalization factor ensuring the
fulfillment of the condition
%нормировочный множитель, обеспечивающий выполнение условия
$\langle\phi_{n+1}|\phi_{n+1}\rangle = 1$), so
that the Hamiltonian in this new basis is tridiagonal \cite{Parlett}:
%В этом новом базисе матрица гамильтониана $\hat{H}$ будет трехдиагональной
%Если построить матрицу гамильтониана в подпространстве, натянутом на вектора $\{ |\phi_{i}\rangle \}_{i=0,1,\dots,n}$, то
%она окажется трехдиагональной:
  \begin{equation}
    T_{n} = \left(
    \begin{array}{ccccc}
     \alpha_0     &  \beta_1      &            &            &            \\
     \beta_1      &  \alpha_1     & \beta_2    &            &            \\
                  &  \beta_2      & \alpha_2   &  \ddots    &            \\
                  &               & \ddots     &  \ddots    &  \beta_n   \\
                  &               &            & \beta_n    &  \alpha_n
    \end{array}                  \right) \ .
  \label{eq:trid}
  \end{equation}
The lowest eigenvalue $E_0^{LT(N)}$ of the matrix $T_{N-1}$ will be called the $N$th approximation ($N=1,2,\dots$) to the
ground state energy $E_0$.
%Наименьшее собственное значение $E_0^{LT(n)}$ матрицы (\ref{eq:trid}) будем называть n-th приближением LT(n) к
%энергии основного состояния $E_0$. При увеличении $n$ происходит постоянное расширение подпространства, натянутого на
With increasing $N$, the expansion of the subspace spanned by the vectors $\{|\phi_{n}\rangle\}_{n=0,..,N-1}$ occurs, so
$E_0^{LT(N)}$ can only decrease:
%вектора  за счет добавления новых базисных
%векторов, при котором энергия основного состояния может только уменьшиться:
$E_0^{LT(1)} \geq E_0^{LT(2)}\geq E_0^{LT(3)}\geq\dots\geq E_0$. Thus, under the condition $\langle \psi_0 |\phi_0\rangle
\neq 0$ Lanczos tridiagonalization gives a monotonically decreasing, guaranteed convergent sequence of approximations to
the true value of the ground state energy \cite{Parlett}.
%Таким образом, при условии $\langle \psi_0 |\phi_0\rangle \neq 0$ Ланцош-тридиагонализация дает монотонно убывающую,
%\emph{гарантированно} сходящуюся последовательность приближений к истинному значению энергии основного состояния $E_0$

%The traditional use of the Lanczos algorithm has been in a purely numerical way, that is to say as a numerical technique
%for exact diagonalisation of very large matrices that arise in treating many-body problems in small finite
%systems\cite{lanczos-D-94}. The potential of taking the Lanczos algorithm far beyond these limitations, into a more
%powerful, universal formalism has not been widely appreciated,
%although some inkling of this was apparent in the suggestion
%of Mattis\cite{trid-M-81} concerning the exact mapping of the many-body problem onto a one-dimensional nearest-neighbour
%model.
%This idea was explored in some applications to the Kondo and Wolff models by Mancini and
%Mattis\cite{trid-MM-83,trid-MM-84,trid-MM-85}. We wish to emphasis to the reader that our approach here is quite different
%from that used in the exact diagonalisation studies of finite systems in following respect - we do not construct a full
%basis for a finite system but manipulate basis vectors and coefficients of an arbitrarily large system analytically and
%symbolically.
%, and we perform every iteration exactly and therefore need not concern ourselves with round-off or loss of orthogonality issues.

The main problem of traditional computational LT is the storage of states $\{ |\phi_{n}\rangle \}$ in the computer memory.
And, typical for many-electron problems, the size of the memory required grows exponentially with the number of electrons.
%Основная проблема При традиционном computational подходе вектора $ |\phi_{0}\rangle $, $|\phi_{1}\rangle$,
%$|\phi_{2}\rangle$ и т.д. хранятся явно в памяти компьютера.
%И typical в многоэлектронных задачах, размер требуемой памяти экспоненциально растет с ростом числа электронов.
Instead, the first few elements of the matrix (\ref{eq:trid}) can be expressed in terms of cumulants
\cite{ManciniMattis1983}, which in many cases can be calculated without storing many-electron states. First, we express
$\alpha_n$ and $\beta_n$ in terms of moments
% Вместо того мы предлагаем использовать полуаналитический метод: несколько первых
%элементов матрицы (\ref{eq:trid}) можно вычислить полуаналитически напрямую (аналитически выразить) через кумулянты, для
%расчета которых во многих случаях можно обойтись без хранения многоэлектронных векторов. Сначала выразим $\alpha_n$ и
%$\beta_n$ их через моменты
$\mu_m$ \cite{WitteHollenberg1994}:
\begin{eqnarray}
\label{eq:alpha_beta}
\alpha_n &=& \frac{\Delta_{n-2}'}{\Delta_{n-1}'}\frac{\Delta_n}{\Delta_{n-1}} + \frac{\Delta_n'}{\Delta_{n-1}'}\frac{\Delta_{n-1}}{\Delta_n} \ , \\
\beta_n^2 &=& \frac{\Delta_n\Delta_{n-2}}{\Delta_{n-1}^2} \nonumber
\end{eqnarray}
where
  \begin{equation} \label{eq:Delta}
    \Delta_n = \left|
    \begin{array}{cccc}
      \mu_0 & \mu_1 & \dots & \mu_n     \\
      \mu_1 & \mu_2 & \dots & \mu_{n+1} \\
      \vdots & \vdots & \ddots & \vdots      \\
      \mu_n & \mu_{n+1} & \dots & \mu_{2n} \\
    \end{array}     \right| ,
  \end{equation}
  \begin{equation} \label{eq:Delta_prime}
    \Delta_n' = \left|
    \begin{array}{cccc}
      \mu_1 & \mu_2 & \dots & \mu_{n+1} \\
      \mu_2 & \mu_3 & \dots & \mu_{n+2} \\
      \vdots & \vdots & \ddots & \vdots      \\
      \mu_{n+1} & \mu_{n+2} & \dots & \mu_{2n+1}
    \end{array}     \right|
  \end{equation}
and $\Delta_{-1}=1, \Delta'_{-2}=0, \Delta'_{-1}=1$. Second, we write the moments in terms of cumulants according to
%Затем моменты выразим через кумулянты согласно соотношению
(\ref{eq:def_cum}). And as a result
%And finally
%В итоге получаются формулы, выражающие матричные элементы трехдиагональной матрицы $T_n$ через кумулянты:
\begin{eqnarray}
\alpha_0 &=& I_1 \ , \nonumber \\
\alpha_1 &=& I_1 + I_3/I_2 \ ,    \nonumber \\
\alpha_2 &=& I_1 + (I_3^3 - 2 I_2 I_3 I_4 + I_2^2 I_5 + 4 I_2^3 I_3) / (-I_2 I_3^2 + I_2^2 I_4 + 2 I_2^4) \ ,  \\
%\alpha_3 &=& I_1 + (72I_2^8I_3 + 36I_2^7I_5 - 4I_2^5(18I_3^3 - I_4I_5 + 6I_3I_6) + 4I_2^6(9I_3I_4 + I_7) -
%I_2^3(264I_3^3I_4 + 7I_4^2I_5 - 20I_3I_5^2 + 8I_3I_4I_6 \nonumber \\
%&+& 4I_3^2I_7) + I_3(15I_3^4I_4 - I_4^4 + 3I_3I_4^2I_5 - I_3^2(I_5^2 + 2I_4I_6) + I_3^3I_7) + 2I_2^4(53I_3I_4^2 + 24I_3^2I_5 - 2I_5I_6 + 2I_4I_7) \\
%&+& I_2I_3(-50I_3^2I_4^2 + 21I_3^3I_5 - 2I_4I_5^2 + 2I_4^2I_6 + 2I_3I_5I_6 - 2I_3I_4I_7) + I_2^2(126I_3^5 + 51I_3I_4^3
%-50I_3^2I_4I_5 + I_5^3 + 12I_3^3I_6 \nonumber \\
%&-& 2I_4I_5I_6 + I_4^2I_7)) / ((2I_2^3 - I_3^2 + I_2I_4)(12I_2^6 - 9I_3^4 + 24I_2^4I_4 - I_4^3 + 2I_3I_4I_5 + I_2^2(7I_4^2
%- 12I_3I_5) - I_3^2I_6 \nonumber \\
%&+& I_2^3(-24I_3^2 + 2I_6) + I_2(12I_3^2I_4 - I_5^2 + I_4I_6))); \nonumber \\
\beta_1^2 &=& I_2 \ , \nonumber \\
\beta_2^2 &=& 2 I_2 + (-I_3^2 + I_2 I_4)/I_2^2 \ . \nonumber
%\beta_3^2 &=&(I_2 (12 I_2^6 - 9 I_3^4 + 24 I_2^4 I_4 - I_4^3 + 2 I_3 I_4 I_5 + I_2^2 (7 I_4^2 - 12 I_3 I_5) - I_3^2 I_6 +
%            I_2^3 (-24 I_3^2 + 2 I_6) +
%            I_2 (12 I_3^2 I_4 - I_5^2 +
%                  I_4 I_6))) \nonumber \\
%                  &/& (2 I_2^3 - I_3^2 + I_2 I_4)^2 . \nonumber
\end{eqnarray}

%Осталось научиться считать кумулянты без хранения многочастичного состояния в памяти компьютера. Это будет сделано ниже
%применительно к конкретным практическим задачам.
It remains to learn how to calculate cumulants without storing the many-electron state in the computer memory. This will be
done below in relation to specific practical problems.
%А кумулянты, как будет показано ниже, можно вычислить без хранения многочастичного вектора в памяти компьютера.

\subsection{Convergence acceleration of a approximations sequence}
It is advisable to accelerate the convergence of the obtained sequence $E_0^{LT(N)}$.
%LT (I) Имеет смысл ускорить сходимость полученной последовательности $LT(N)$. Попробуем воспользоваться наиболее популярным The
Let's apply the most popular $\varepsilon$-algorithm \cite{Brezinski}. Let $S_n$ be some initial sequence ($n =
0,1,\dots$).
%Пусть $S_n$ -- исходная последовательность ($n = 0,1,\dots$).
For $k=0,1,\dots$ we construct new sequences
\begin{equation}
\varepsilon_{k+1}^{(n)} = \varepsilon_{k-1}^{(n+1)} + \frac{1}{\varepsilon_k^{(n+1)} - \varepsilon_k^{(n)}} \ ,
\end{equation}
where $\varepsilon_0^{(n)} = S_n$, $\varepsilon_{-1}^{(n)} = 0$. Then, for a wide class of sequences $S_n$, it is true that
the rate of convergence of sequences $\varepsilon_{2k}^{(n)}$ is the faster, the larger $k$.
%Тогда для широкого класса последовательностей $S_n$ верно, что скорости сходимости последовательностей
%$\varepsilon_{2k}^{(n)}$ %with $k= 1,2,...$
%тем выше, чем больше $k$.
%имеют более быструю сходимость, чем исходная $S_n$.

Below, when applying this method, we will assume $S_n = E_0^{LT(n+1)}$.

\section{Applications}
\subsection{Anharmonic oscillator}
Let us apply the described method to calculate the ground state energy of the anharmonic
oscillator
%Применим описанную методику для вычисления энергии основного состояния ангармонического осциллятора:
\begin{equation}
\left(-\frac{1}{2}\frac{d^2}{dx^2} + \frac{x^2}{2} + g x^4\right)|\psi_0\rangle = E_0|\psi_0\rangle \ ,
\end{equation}
where, as is well known, the expansion in %(terms of the coupling constant)
$g$ gives a diverging series \cite{BenderWu1969}.
%где хорошо известно, что формальное разложение по $g$ дает для $E_0$ расходящийся ряд.

%$$E_0 = 1/2$$
In this case, the moments (\ref{eq:mu_n}) can be calculated simply by the $n$-fold action of the Hamilton operator on the
initial state $|\phi_0\rangle = \frac{e^{-x^2/2}}{\pi^{1/4}}$ followed by integration over $x$.
%In this case, the moments
%(\ref{eq:mu_n}) can be calculated simply by acting with the Hamilton operator on the seed vector.
%В данном случае
%моменты $\mu_n$ можно получить просто действуя оператором Гамильтона на затравочный вектор $|\phi_0\rangle =
%\frac{e^{-x^2/2}}{\pi^{1/4}}$.
Then $\alpha_n$ and $\beta_n$ are calculated using formulas
%Затем $\alpha_n$ и $\beta_n$ вычисляются по формулам
(\ref{eq:alpha_beta}), (\ref{eq:Delta}) and (\ref{eq:Delta_prime}).
%ангармонизм $V = g x^4$
%$$\mu_1 = \frac{1}{2} + \frac{3}{4}g$$
%$$\mu_2 = \frac{1}{4} + \frac{3}{4}g + \frac{105}{16}g^2$$
\begin{table}[htb]
\caption{The ground state energy $E_0$ of the anharmonic oscillator at different $g$}
\begin{tabular}{|c|c|c|c|c|c|}
\hline
$g$& 0.1 & 0.3 & 0.5 & 1 & 2        \\
\hline
%Pad\'e(g)[1/1] & 0.555556 & 0.609756 & 0.636364 & 0.666667 & 0.6875 \\
%Pad\'e(g)[2/2] & 0.55877  & 0.630408 & 0.674145 & 0.733845 & 0.782913 \\
%Pad\'e(g)[3/3] & 0.559092 & 0.635536 & 0.6869   & 0.765063 & 0.838063 \\
$LT(1)=I_1$          & 0.575    & 0.725    & 0.875 & 1.25 & 2\\
CMX(2)               & 0.562940 & 0.670592 & 0.774160 & 1.029817 & 1.53846  \\
CMX(3)               & 0.560608 & 0.655804 & 0.743471 & 0.955635 & 1.37375  \\
$D$-Pad\'e$[0/2]$    & 0.552529 & 0.603212 & 0.622528 & 0.601112 & 0.40048     \\
$D$-Pad\'e$[0/3]$    & 0.557691 & 0.640572 & 0.716298 & 0.901608 & 1.26980   \\
LT(2)                & 0.562969 & 0.670887 & 0.774834 & 1.031567 & 1.54249  \\
LT(3)                & 0.560621 & 0.656031 & 0.744094 & 0.957527 & 1.37849  \\
LT(4)                & 0.559802 & 0.649535 & 0.729617 & 0.920075 & 1.29192  \\
LT(5)                & 0.559459 & 0.646010 & 0.721307 & 0.897360 & 1.23757 \\
LT(6)                & 0.559304 & 0.643851 & 0.715955 & 0.882059 & 1.19991\\
LT(7)                & 0.559230 & 0.642423 & 0.712242 & 0.871020 & 1.17206\\
$\varepsilon_2^{(0)}$& 0.560051 & 0.650410 & 0.730482 & 0.919561 & 1.28687 \\
$\varepsilon_2^{(1)}$& 0.559364 & 0.644486 & 0.716732 & 0.881740 & 1.19509 \\
$\varepsilon_2^{(2)}$& 0.559212 & 0.641831 & 0.710105 & 0.862348 & 1.14596 \\
$\varepsilon_2^{(3)}$& 0.559176 & 0.640434 & 0.706275 & 0.850479 & 1.11484 \\
$\varepsilon_2^{(4)}$& 0.559162 & 0.639632 & 0.703826 & 0.842436 & 1.09313 \\
$\varepsilon_4^{(0)}$& 0.559195 & 0.641091 & 0.708015 & 0.855645 & 1.12802 \\
$\varepsilon_4^{(1)}$& 0.559168 & 0.639698 & 0.703911 & 0.842350 & 1.09229 \\
$\varepsilon_4^{(2)}$& 0.559156 & 0.639027 & 0.701596 & 0.834198 & 1.06954 \\
$\varepsilon_6^{(0)}$& 0.559152 & 0.638893 & 0.701052 & 0.832090 & 1.06339 \\
%Aitken-2a            & 0.559168 & 0.639677 & 0.703941 & 0.841943 & 1.08936 \\
%Aitken-2b            & 0.559165 & 0.638881 & 0.701029 & 0.831752 & 1.06108 \\
%Aitken-2c            & 0.559154 & 0.638553 & 0.69948  & 0.825537 & 1.04304 \\
%Aitken-3             & 0.559169 & 0.638323 & 0.697719 & 0.815824 & 1.01124 \\
\hline
exact\cite{VinetteCizek1991}& 0.559146 & 0.637992 & 0.696176 & 0.803771 & 0.95157       \\
\hline
%CMX(2)&0.5629396984924623 & 0.6705919395465995 & 0.774159663865546 & 1.029816513761468 & 1.538461538461539  \\
%CMX(3)&0.5606081087935313 & 0.6558040810196106 & 0.743470980711108 & 0.9556347626040926 & 1.373747137164597  \\
%$D$-Pad\'e$[0/2]$&0.552529070005 & 0.603212014929 & 0.622527536594 & 0.601112273254 & 0.400476775132     \\
%$D$-Pad\'e$[0/3]$&0.557690522328 & 0.640572245761 & 0.71629803896 & 0.901607963304 & 1.26979932106   \\
%LT(2)& 0.5629687939535542 & 0.6708869742624157 & 0.7748341643450392 & 1.031567441562768 & 1.542486889354094  \\
%LT(3)& 0.5606205425576836 & 0.6560314740567039 & 0.7440937459780969 & 0.957526783602996 & 1.378494525379869  \\
%LT(4)& 0.5598021081331687 & 0.6495348284745691 & 0.729617307692031 & 0.9200749053219813 & 1.291915658645669  \\
%LT(5)& 0.5594591101375173 & 0.6460104173872442 & 0.721306684261695 & 0.8973599090877434 & 1.237571745837488 \\
%LT(6)& 0.5593039889488137 & 0.643850875895555  & 0.7159549140984091& 0.8820586778406767 & 1.199905544102715\\
%LT(7)& 0.559229943087456  & 0.6424225197597484 & 0.7122415659733542& 0.8710202480113777 & 1.172061492768789\\
\end{tabular}
\label{tabl:osc}
\end{table}
As expected, LT gives a monotonically converging sequence of approximations for all $g$.
%Как и следовало ожидать, SALT дает монотонно сходящуюся последовательность приближений для всех значений $g$.
The convergence-accelerating $\varepsilon$-algorithm significantly improves the existing results.
%Ускоряющий метод $\varepsilon$-algorithm существенно улучшает имеющиеся результаты.

\subsection{Correlation energy in a molecule}

One of the first attempts to use cumulant expansions in quantum chemistry has been made in \cite{Cioslowski}.
% Одной из первых попыток использовать кумулянтное разложение в квантовой химии была работа \cite{Cioslowski}.
%В этой работе ставился
%(и получил утвердительный ответ)
%вопрос: насколько точно можно оценить $E_0$, если кумулянты уже откуда-либо известны.
%например, из полноконфигурационного расчета.
%Там вычислялись моменты по формулам типа
In this work, the cumulants were calculated from the appropriated traces of matrices representing powers of Hamiltonian.
% using formulas of the type
%In this paper points were calculated using the formulas of type
%\begin{eqnarray}
%\label{eq:mom_direct} \mu_2 &=& \langle\phi_0| \hat{H}^2 |\phi_0\rangle = \sum_i \langle\phi_0| \hat{H} |\phi_i\rangle
%\langle\phi_i| \hat{H}|\phi_0\rangle , \\ \nonumber \mu_3 &=& \langle\phi_0| \hat{H}^3 |\phi_0\rangle = \sum_{ij}
%\langle\phi_0| \hat{H} |\phi_i\rangle \langle\phi_i| \hat{H}|\phi_j\rangle \langle\phi_j| \hat{H}|\phi_0\rangle,
%\end{eqnarray}
%and then cumulants according to (\ref{eq:def_cum}). %а затем кумулянты согласно (\ref{cumulants}).
%This method requires a search of all configurations, so as expensive as the FCI.
%This method requires enumeration of all configurations, therefore it is as expensive as FCI.
This method requires the use of all many-electron configurations, so as expensive as the FCI.
%This method requires iterating over all configurations is therefore as expensive as FCI.
%Этот метод требует перебора всех конфигураций, поэтому столь же дорог, как и FCI.
%Однако для практических целей кумулянты нужно вычислять более экономичными, чем FCI, методами.
%However, for practical purposes cumulants need to calculate a cost-effective way.
But for practical purposes, the cumulants need to be calculated in more economical way.
%Такое вычисление (по технике спариваний) кумулянтов до третьего порядка было проведено в \cite{CiosPoir1987}.
%Such a calculation (by the technique of pairing) to the third order cumulant was conducted in \ cite {CiosPoir1987}.
Such a calculation (by pairing technique) of cumulants up to third order was carried out in \cite{CiosPoir1987}, but using
CMX(2) only about half of the correlation energy was captured.

In this work, we have derived formulas for cumulants up to the fifth order using the technique described below.

Suppose that the Hartree--Fock equations for a molecule are written and solved by standard methods. Let us write the
Hamiltonian of the problem in the second quantization representation \cite{Jorgensen} using solutions of the restricted
Hartree--Fock (RHF) equations $\chi_i(\textbf{r})$ as single-electron states:
%Положим, что уравнения Хартри--Фока для молекулы записаны и решены стандартными методами. Запишем гамильтониан задачи в
%представлении вторичного квантования \cite{Jorgensen}, используя в качестве одноэлектронных состояний вектора $\psi_i(1)$,
%являющиеся решениями уравнений Хартри--Фока:
%\begin{eqnarray}
%\hat{H} &=& \hat{T} - \hat{V} + \hat{W}  \ , \nonumber \\
%\hat{T} &=& \sum_{i\sigma} \varepsilon_i c_{i\sigma}^{\dagger }c_{i\sigma} \ , \\
%\hat{V} &=& \sum_\gamma^{E_F}
%\sum_{ij\sigma} (2[ij|\gamma\gamma] - [i\gamma|\gamma j]) c_{i\sigma}^{\dagger }c_{j\sigma} \ , \nonumber \\
%\hat{W} &=& \frac{1}{2}\sum_{ijkl\sigma\sigma'} [ij|kl] c_{i\sigma}^{\dagger }c_{k\sigma'}^{\dagger} c_{l\sigma'}
%c_{j\sigma} \ , \nonumber \label{eq:Ham}
%\end{eqnarray}
\begin{equation}
\hat{H} = \sum_{i\sigma} \varepsilon_i c_{i\sigma}^{\dagger }c_{i\sigma} - \sum_\gamma \sum_{ij\sigma} (2[ij|\gamma\gamma]
- [i\gamma|\gamma j]) c_{i\sigma}^{\dagger }c_{j\sigma} +
 \frac{1}{2}\sum_{ijkl\sigma\sigma'} [ij|kl] c_{i\sigma}^{\dagger }c_{k\sigma'}^{\dagger} c_{l\sigma'} c_{j\sigma} \ ,
\label{eq:Ham}
\end{equation}
where summation over $\gamma$ is carried out over occupied orbitals, $\varepsilon_i$ -- the energy of the \emph{i}th
orbital and Coulomb integrals
\begin{equation}
[ij|kl] \equiv \int\chi_i^*(\textbf{r}_1)\chi_j(\textbf{r}_1) \frac{1}{r_{12}} \chi_k^*(\textbf{r}_2)\chi_l(\textbf{r}_2) d\textbf{r}_1 d\textbf{r}_2 \ . %??? dr_1 dr_2
\end{equation}

For the initial ket $|\phi_0\rangle$ being the RHF wavefunction, $I_1$ is the Hartree--Fock energy, and other cumulants
%many-operator average included in cumulant $I_m$
can be calculated using Wick's pairing technique (for more details see \cite{Zhuravlev2016,Zhuravlev2020}).
%Выберем затравочную волновую функцию $|\phi_0\rangle$ в виде детерминанта Слэтера
%\begin{eqnarray}
%|\phi_0\rangle = \prod_{l} c_{l}^\dagger |0\rangle
%\end{eqnarray}
%($|0\rangle$ -- состояние без электронов, $l$ обозначает занятые орбитали). Рассмотрим гамильтониан, описывающий электроны
%в молекуле. В качестве затравочного состояния $|\phi_0\rangle$ возьмем волновую функцию, полученную ограниченным методом
%Хартри--Фока. Тогда многооператорные средние, входящие в кумулянт $I_m$, могут быть вычислены с помощью техники спариваний \cite{Zhuravlev2016}.
%Каждое такое среднее can be computed easily, but there are too many of them to perform all the calculations manually.
%To complete the task the symbolic manipulation computer program was written that performs these calculations.
%В итоге кумулянт  , использовав технику спариваний и выполнив необходимые аналитические вычисления с помощью специально
%написанной компьютерной программы, получаем окончательные выражения для кумулянтов:
Performing necessary analytical calculations%by the symbolic computer program
, we have obtained the final expressions for
cumulants:
\begin{eqnarray}
\label{eq:I_n}
I_2 &=& \sum_{ab r^*s^*} [ar|bs]\cdot(2[ar|bs] - [as|br]) \ , \nonumber \\
I_3 &=& \sum_{ab r^*s^*} 2[ar|bs]\cdot(2[ar|bs]-[as|br])\cdot(\varepsilon_r - \varepsilon_a) \nonumber \\
&+& \sum_{abc r^*s^*t^*} 2[ar|bs]\cdot\{2[ar|ct] (2[bs|ct]-[bt|cs]) - [br|ct](2[as|ct]\\
           &-& [at|cs]) - [rt|ca](2[sb|tc]-[sc|tb]) - [rt|cb](2[sc|ta] -[sa|tc])\} \nonumber \\
&+& \sum_{ab r^*s^*t^*u^*} [ar|bs][rt|su](2[at|bu] - [au|bt]) \nonumber \\
&+& \sum_{abcd r^*s^*} [ar|bs][ac|bd](2[cr|ds] - [dr|cs]) \ ,  \nonumber
\end{eqnarray}
where the starred (unstarred) indices refer to virtual (occupied) orbitals.
%где наличие звездочки около индекса означает, что суммирование по этому индексу производится по пустым орбиталям, а её отсутствие -- по заполненным.
%Следует сразу отметить важную вещь: выведенные формулы
%отличаются от приведенных в \cite{CiosPoir1987}: слагаемое с суммой по $abcr^*s^*t^*$ другое, при том, что остальные
%слагаемые совпадают с вычисленными в указанной работе.
It should be noted that the derived formulas differ from those given in \cite{CiosPoir1987}: the term with the sum over
$abcr^*s^*t^*$ is different, while the rest terms coincide with those calculated in the indicated work.
%По-видимому, в работе \cite{CiosPoir1987} была совершена ошибка при
%выводе формул.
Apparently, an error was made in this article when deriving the formula for $I_3$. At the same time, the results of
calculations using our formulas (\ref{eq:I_n}) completely coincide with the results obtained by direct computation of the
traces of the Hamiltonian power matrices.

%(\ref{eq:mom_direct}) and (\ref{eq:def_cum}).
% Расчеты по формулам (\ref{eq:I_n}) полностью совпадают с расчетами по прямым формулам (\ref{eq:cum_direct}).

The explicit formulas for cumulants $I_4$ and $I_5$ were obtained using a program for analytical calculations, but are not
given here because of their large size (in particular, the expression for $I_4$ contains more than two hundred lines).
As an illustration, the correlation energy of a hydrogen molecule was calculated within the 6-31** basis. The
results of the ground state energy calculations using the described technique are presented in Table \ref{tabl:H2}.
% В качестве базисных использовались гауссовы орбитали 6-31G**. Результаты вычисления
%энергии основного состояния приведены в таблице \ref{tabl:H2}. As one can see the estimation for the ground state energy
%converges to its exact value with increasing of the number of the cumulants known.
%The new AD-method radically accelerates
%the convergence rate compared to $D$-Pad\'e approximation for large $v/w$, that is, where there is a large gap $\Delta$ in
%the energy spectrum, despite the fact that index $\alpha$ is only a rough estimate for $\Delta$. For $v\lesssim w$ the
%accuracy of the AD-method are comparable to the accuracy of the $D$-Pad\'e method.
\begin{table}[htb]
\caption{Ground state energy (hartrees) for the hydrogen molecule with 1.4 a.u. H--H distance; third column: captured part
of the correlation energy}
\begin{tabular}{|c|c|c|c|c|c|}
\hline
%$R$ & 1.4 &        \\
HF   & -1.131287 &  \\ %TT
\hline
MP2   & -1.157629  & 77.8\% \\
MP3   & -1.162488  & 92.1\% \\
%\hline
%CMX(2) acc. \cite{CiosPoir1987}   & -1.148109765735355  & 49.7\% \\
CMX(2)             & -1.156932 & 75.7\%    \\
CMX(3)             & -1.162512 & 92.2\%   \\
$D$-Pad\'e$[0/2]$  & -1.175371 & 130.1\%   \\
$D$-Pad\'e$[0/3]$  & -1.166466 & 103.9\%  \\
LT(2)              & -1.156769 & 75.2\% \\
LT(3)              & -1.162338 & 91.7\% \\
%Lanczos($T_3$)&  -1.16433266304969 (0.975607) \\
$\varepsilon_2^{(0)}$&  -1.163895  & 96.3\%\\
%$\varepsilon_2^{(1)}$&  -1.165445639231638 (1.00847) \\
\hline
FCI  & -1.165159      &         \\
\hline
\end{tabular}
\label{tabl:H2}
\end{table}

Thus, cumulant Lanczos tridiagonalization, being comparable in terms of resource requirements with the M{\o}ller--Plesset
perturbation theory, yield a guaranteed convergent, monotonic sequence of approximations to the correlation energy. I would
like to draw the attention of researchers to this method.
%Таким образом, мы получили несколько первых членов монотонно сходящейся последовательности поправок к методу Хартри--Фока,
%причем требования к размерам памяти здесь такие же, как и у ХФ.

The research was carried out within the state assignment of the Ministry Science and Higher Education of the Russian
Federation (theme ``Quantum'' no. AAAA-A18-118020190095-4).

\section*{References}
\bibliography{mybibfile}
\end{document}